\documentclass[preprint,showpacs,preprintnumbers,amsmath,amssymb]{revtex4}

\usepackage{graphicx}
\usepackage{dcolumn}
\usepackage{bm}

\begin{document}

\preprint{APS/123-QED}

\title{Roles of two successive phase transitions in new spin-Peierls system TiOBr}

\author{T. Sasaki,$^1$ M. Mizumaki,$^2$ T. Nagai,$^3$ T. Asaka,$^3$ K. Kato,$^{2,4}$ M. Takata,$^{2,4}$ Y. Matsui,$^3$ and J. Akimitsu$^1$
}
\affiliation{$^1$Department of physics, Aoyama-Gakuin University, Sagamihara, Kanagawa 229-8558\\
$^2$Japan Synchrotron Radiation Reserch Institute (JASRI), Spring-8, Hyogo 679-5198\\
$^3$High Voltage Electron Microscopy Station, National Institute for Materials Science, Tsukuba 305-0044\\
$^4$CREST, Japan Science and Technology Corporation (JST), Kawaguchi, Saitama 332-0012}

\date{\today}

\begin{abstract}
In this sturdy, we determine the roles of two successive phase transitions in the new spin-Peierls system TiOBr by electron and synchrotron X-ray diffraction analyses. Results show an incommensurate superstructure along the $h$- and $k$-directions between $T_{\textrm c1}$=27K and $T_{\textrm c2}$=47K, and a twofold superstructure which is related to a spin-Peierls lattice distortion below $T_{\textrm c1}$. The diffuse scattering observed above $T_{\textrm c2}$ indicates that a structural correlation develops at a high temperature. We conclude that $T_{\textrm c2}$ is a second-order lock-in temperature, which is related to the spin-Peierls lattice distortion with the incommensurate structure, and that $T_{\textrm c1}$ is from incommensurate to commensurate phase transition temperature accompanying the first-order spin-Peierls lattice distortion.
\end{abstract}

\pacs{Valid PACS appear here}
\maketitle

 Low-dimensional $S$=1/2 quantum spin systems have several unique features due to large quantum spin fluctuations. For instance, a one-dimensional (1D) chain coupled to three-dimensional lattice system undergoes spin-Peierls transition to a nonmagnetic dimerized state. This exotic phenomenon might be able to help us understand the roles of lattice and spin degrees of freedom in quasi-low-dimensional spin systems. Actually, detailed studies of the first inorganic spin-Peierls compound CuGeO$_3$ have deepened the understanding of a spin-Peierls system, and opened higher level physics \cite{1,2}.

 Recently, Seidel $et$ $al$. have suggested that TiO$X$ ($X$=Cl, Br) is a spin-Peierls compound with on orbital degree of freedom \cite{Seidel}. TiO$X$ has a FeOCl-type crystal structure \cite{Beynon,Sasaki}, where the TiO$_4X_2$ bilayers separate from each other along the $c$-axis. The chains are composed of edge-shared TiO$_4X_2$ octahedra in a direct contact with each other along the $b$-axis due to the occupation of the $d_{xy}$ orbital in the ground state \cite{Seidel,Kataev}. It is still an open question whether the single valence of Ti$^{3+}$ has occupied its orbital degree of freedom in this system \cite{Kataev,Ruckamp}. The evidence indicating that TiO$X$ belongs to a new spin-Peierls system has been confirmed by the susceptibility \cite{Seidel, Kato}, nuclear magnetic resonance (NMR) \cite{Imai,Kikuchi} and X-ray diffraction \cite{Sasaki,Shaz,Palatinus} measurements.

TiOBr exhibits two successive phase transitions with $T_{\textrm c1}$=27K and $T_{\textrm c2}$=47K. The temperature dependence of susceptibility starts to gradually decrease at $T_{\textrm c2}$ and show a sudden drop to zero at $T_{\textrm c1}$. These transitions have been reported to be of the first-order at $T_{\textrm c1}$ and second-order transition at $T_{\textrm c2}$ by NMR \cite{Imai} and heat capacity measurements \cite{Hemberger}. To elucidate the origin of these transitions, we have performed electron and synchrotron radiation (SR) X-ray diffraction analyses. Below $T_{\textrm c1}$, superlattice reflections were observed at ($h$, $k+1/2$, $l$) positions, which are related to the spin-Peierls lattice distortion \cite{Sasaki,Shaz}. If $T_{\textrm c1}$ is the spin-Peierls transition temperature, TiO$X$ is a new spin-Peierls compound having the first-order phase transition. Although the origin of the two successive phase transitions has not yet been clarified, Imai and Chou \cite{Imai} suggested that "the emergence of the spin gap at $T_{\textrm c2}$ is accompanied by an orbital-order (possibly incommensurate)". Above $T_{\textrm c1}$, we found the incommensurate superlattice reflections by SR X-ray and electron diffraction analyses \cite{SCES}. Recently, these results have independently been confirmed by other research groups \cite{Ruckamp,Smaalen}. In this paper, we discuss the key roles of two successive phase transitions.


Polycrystalline and single crystal of TiOBr samples were prepared by a chemical vapor transport technique \cite{Sasaki,Kato}. SR X-ray diffraction experiments were carried out using a four-circle diffractometer at BL46XU, SPring-8. The X-ray wavelength determined using Si (111) monochromators was 1.0332\AA. A single crystal with dimensions of $1.5 \times 6 \times 0.03$mm$^3$ was glued on a BN plate, which was mounted on a refrigerator with the $(hk0)$ reciprocal plane perpendicular to the $\chi$-axis. At electron diffraction analysis, a single crystal was thinned bu Ar$^+$ ion sputtering, and using a Hitachi HF-3000S high-voltage TEM operating at 300kV at the High Voltage Electron Microscopy Station, NIMS.


We examined the electron diffraction patterns at the $(hk0)$ reciprocal lattice plane at various temperatures in the single crystal of TiOBr. Figure 1 shows the electron diffraction patterns of TiOBr at the $(hk0)$ plane. At the $(hk0)$ plane, where $h$ and $k$ are integers, strong reflections are indexed as fundamental reflections of orthorhombic Pmmn. However, weak spots are observed at $h+k=2n+1$, which are the forbidden reflections of orthorhombic $Pmmn$. The crystal symmetry of TiOBr may be different from that of orthorhombic $Pmmn$, because the forbidden reflections are observed even at room temperature \cite{STAM}. Below $T_{\textrm c1}$ (commensurate phase), the electron diffraction patterns show twofold superstructure along the $k$-direction at ($h$, $k+1/2$, 0), which is in good agreement with the X-ray diffraction results \cite{Sasaki,Shaz}. The superlattice reflections split into two spots along the $h$-direction with diffuse scattering and are out of $k=1/2+n$ along the $k$-direction (see Fig.1 (b) and (d)$\sim$(i)) above $T_{\textrm c1}$. To confirm the electron diffraction results, we performed an SR single crystal X-ray diffraction analysis. As shown in Fig.2, the twofold superstructure is observed below $T_{\textrm c1}$, and an incommensurate structure is produced along the $h$- and $k$-directions between $T_{\textrm c1}$ and $T_{\textrm c2}$ (intermediate region).

Figure 3 shows the temperature dependences of superlattice reflections around (0 2.5 0) determined by electron (open circle) and SR X-ray (closed circle) diffraction analyses. The electron diffraction result does not agree well with the X-ray diffraction result in the intermediate region. This is ascribed to the difference in sensitivity to atoms between the electron and X-ray probes. In addition, as the sample is locally heated by the electron diffraction technique, temperature is unstable around the observed area. Therefore, it is difficult to observe clearly separated incommensurate double peaks along the $h$-direction in electron diffraction, although the clear incommensurate peaks are observed in SR X-ray diffraction (compare Fig.1 and Fig.2). The remarkable behaviors of $\delta_h$ and $\delta_k$  at $T_{\textrm c1}$ and $T_{\textrm c2}$ are shown in Fig.3; $\delta_h$ continuously changs only within the intermediate region and becomes constant (0.083[r.l.u.]) above $T_{\textrm c2}$. $\delta_k$ jumps from 0.5 to 0.495 at $T_{\textrm c1}$ with increasing temperature, and is inversely proportional to temperature above $T_{\textrm c1}$, which is expressed as the function $\delta_k \propto -4.5 \times 10^{-4}T$. The jumps of $\delta_h$ and $\delta_k$ at $T_{\textrm c1}$ imply a first-order incommensurate (above $T_{\textrm c1}$) to commensurate (below $T_{\textrm c1}$) phase transition. However, $\delta_k$ is unaffected by the phase transition at $T_{\textrm c2}$; in other words, $\delta_h$ and $\delta_k$ behave independently at $T_{\textrm c2}$. This result implies that each temperature has a unique essential role in spin-Peierls transition


Figure 4 shows the temperature dependences of the integrated intensities of superlattice reflections measured by $\omega$-scan at around (0 2.5 0). The two integrated intensities of the incommensurate reflections at ($\pm \delta_h$, $2+\delta_k$, 0) are similar in this experiment \cite{Comment}, and gradually decrease to $T_{\textrm c2}$. This implies that $T_{\textrm c2}$ is a second order phase transition temperature. The dashed line in Fig.4 is a fitting result within the intermediate region at ($-\delta_h$, $2+\delta_k$, 0) obtained using the function $I_{inco}(T) \propto (1-T/T_{\textrm c2})^{2 \beta}$  with $T_{\textrm c2}=47.0 \pm 0.1$K and $2\beta=0.193\pm0.071$ \cite{fit1}. $\beta$ is the exponent obtained from the temperature dependence of the atomic displacement d since $I \propto d^2$. The fitting result demonstrates that $\beta$ of TiOBr does not agree with the value ($2\beta=0.66$) of CuGeO$_3$. The integrated intensities tend to decrease at around $T_{\textrm c1}$. This is probably due to the fact that we could not measure the intensities accurately because the superlattice reflections are always accompanied by the diffuse scattering along the $h$-direction in the vicinity of $T_{\textrm c1}$ (see Fig.2 at 25K).


To measure the critical fluctuations at $T_{\textrm c1}$ and $T_{\textrm c2}$, we chose the strongest superlattice peak at ($-2+\delta_h$, $3+\delta_k$, $1$). The correlation lengths, $\xi$, of the critical fluctuations were determined from the reciprocal of half width at half maximum, HWHM, which was corrected on the basis of the experimental resolution. Figure 5 shows the temperature dependences of HWHM along the $h$-, $k$- and $l$-directions. HWHM behaves as $\sqrt{(T-T_{\textrm c2})}$ with the average value of $T_{\textrm c2}=48.7\pm1.5$K. The diffuse scattering along the $h$- and $l$-directions seems to be wider than that along the $k$-direction at $T_{\textrm c1}$ and above $T_{\textrm c2}$. $\xi_k$ decreases very slowly with increasing temperature, such as $\xi_k=272$\AA ($=78\times b$) at 51K. The anisotropy ratio of correlation lengths at 51K is $\xi_h: \xi_k: \xi_l \sim 1.5:13:1$. This implies that the correlation length along chains increases at a high temperature (for example, $\xi_k < b=13158$K) in this system.

Our present study unambiguously  shows that the superlattice reflections split along the $h$- and $k$-directions from the twofold superstructures along the $h$-direction at ($h$, $k+1/2$, 0) above $T_{\textrm c1}$. The incommensurate structures along the $h$- and $k$-directions behave of each other independently. We believe that clarifying the origin of incommensurate structure is the key to understanding the TiO$X$ system.

Recently, R\"{u}ckamp $et$ $al$. and van Smaalen $et$ $al$. have suggested the crystal structure in the incommensurate phase for TiOBr \cite{Ruckamp,Smaalen}; however, they could not experimentally determine whether the final structure is monoclinic or orthorhombic (see Fig.3 of van Smaalen $et$ $al$.), because they did not have the direct information on the crystal symmetry. Our electron diffraction result provides more information necessary for discussing for the crystal symmetry. The incommensurate peaks are different between Fig.1 (e) and (f), which implies that there are two domains perpendicular to the chain direction at the $(hk0)$ plane, as confirmed by selected-area electron diffraction. More detailed experiments for crystal symmetry determination are now in progress \cite{STAM}. In SR X-ray diffraction analysis, the integrated intensities of both the commensurate at (0 2.5 0) and the incommensurate (total) are scaled using the universal function (the solid line in Fig.4), which indicates that the relationship between the commensurate and incommensurate phases is not independent of roles. The integrated intensity of the superlattice reflection is directly related to the length of the lattice distortions, for which energy may be stored in the vicinity of $T_{\textrm c1}$.


The lattice distortion with the incommensurate structure occurs at $T_{\textrm c2}$. The twofold superlattice reflections confirm the presence of the (commensurate) spin-Peierls lattice distortion below $T_{\textrm c1}$, at which the first order incommensurate to commensurate phase transition occurs. Here, we indicate the phenomena occurring at $T_{\textrm c1}$ and $T_{\textrm c2}$, and the roles of the two successive phase transition

As shown in Fig.3, $\delta_k$ is not affected by the lattice distortion with the incommensurate structure at $T_{\textrm c2}$, and the anisotropic diffuse scattering is produced above $T_{\textrm c2}$ up to 150K, which corresponds to the results of ESR and  infrared optical properties \cite{Kataev,Caimi,Lemmens}. With decreasing temperature down to $T_{\textrm c2}$, the structural short-ranged correlation length increases. Although the structural distortion becomes long-ranged at $T_{\textrm c2}$, no twofold distortion is observed in the intermediate region. Since the incommensurate superstructure along the $k$-direction is dominated by the strength of interchain interaction, the unaffected $\delta_k$ at $T_{\textrm c2}$ clearly shows that the interaction within chains is dominant in this material. As shown in Fig.5, the strong diffuse scattering above $T_{\textrm c2}$ is observed, which clearly indicates the local lattice distortion due to the spin-Peierls instability.

As shown in Fig.3, the incommensurability along the h-direction, $\delta_h$ (=0.083), is almost equal to the twelvefold unit cell, which approximately corresponds to the correlation length that relaxes the frustration between chains. The displacement of one atom strongly influences the neighboring atoms. Below 150K, the local incommensurate structure is caused by  the short-ranged distortion within chains, and exhibits on interchain interactions with twelvefold correlation lengths. This is considered to be the key role of the diffuse scattering with the incommensurate structure. With decreasing temperature, the long-ranged distortion with the incommensurate structure is locked in at $T_{\textrm c2}$. Between $T_{\textrm c1}$ and $T_{\textrm c2}$, $\delta_k$  increases toward 1/2. This is because the intrachain interaction (commensurability energy) overcomes the interchain interaction (incommensurability energy), and the commensurability accompanying the first-order transition is established.

In summary, the incommensurate to commensurate phase transition was clearly observed by electron and SR x-ray diffraction analyses in the single crystal of TiOBr. Superlattice reflections were observed at ($h$, $k+1/2$, $l$) below $T_{\textrm c1}$ and at ($h \pm \delta _h$, $k+\delta_k$, $l$); $h,k,l$=integer and $h+k=2n$ for $(hk0)$, above $T_{\textrm c1}$.  The diffuse scattering was clearly observed along the $h$- and $l$-directions at $T_{\textrm c1}$ and above $T_{\textrm c2}$, indicating the development of the structural correlation along chains far above $T_{\textrm c2}$. This suggests that $T_{\textrm c1}$ is a first-order incommensurate to commensurate phase transition temperature with the spin-Peierls lattice distortion, and that $T_{\textrm c2}$ is a second order lock-in temperature, which is related to the spin-Peierls lattice distortion with the incommensurate structure


\begin{acknowledgments}
We would like to thank H. Sawa and T. Yokoo (KEK) for having interest in this work and for fruitful discussion. The Aoyama-Gakuin group was partly supported by the 21$^{st}$ COE program. The synchrotron radiation experiment was performed at BL46XU (R04B46XU-0020N) in SPring-8 with the approval of the Japan Synchrotron Radiation Research Institute (JASRI). A part of the electron diffraction experiment was supported by the "Nanotechnology Support Project" of the Ministry of Education, Culture, Sports, Science and Technology, MEXT, Japan
\end{acknowledgments}


\newpage
{\bf Figure captions}\\
\\
Figure 1 Electron diffraction patterns of TiOBr at $(hk0)$ plane. (a) Commensurate phase, (b) incommensurate (intermediate) phase, (c) high-temperature phase, and (d)$\sim$(i) temperature dependence of superlattice reflections around (040).\\

Figure 2 SR X-ray diffraction mesh scans around (0 2.5 0) on $(hk0)$ plane at each temperature. Intensities are normalized using monitor counts and shown on the log scale. \\

Figure 3 Temperature dependences of (a) $\delta_h$ and (b) $\delta_k$ at ($-\delta_h$, $2+\delta_k$, 0) determined by electron (open circle) and SR X-ray (closed circle) diffractions analyses.\\

Figure 4 Temperature dependences of integrated intensities of superlattice reflections measured by $\omega$-scan around (0 2.5 0). We measured the reflections at two incommensurate positions ($-\delta_h$, $2+\delta_k$, 0) (open square) and ($\delta_h$, $2+\delta_k$, 0)  (closed triangle) above $T_{\textrm c1}$, and at a commensurate position (0 2.5 0) (open circle) below $T_{\textrm c1}$. Closed circles indicate the total intensities of the two incommensurate intensities. The inset shows the rocking curves obtained at (0 2.5 0) 7.5K (closed square), (0.0773, 0.4897, 0) 32.5K (open triangle) and (0.083, 0.4838, 0) 49K (cross).\\

Figure 5 Temperature dependences of HWHM of superlattice peaks at ($-2-\delta_h$, $3+\delta_k$, 1) along $h$-(open circle), $k$-(closed square) and $l$-(closed triangle) directions. The lines indicate the best fits of these data to a $\sqrt{(T-T_{\textrm c2})}$ law. The inset shows the diffuse scatterings at 45K (closed circle), 50K (open square) and 57.5K (cross) along the $h$-direction.\\

\newpage

\begin{figure}[htbp]
\includegraphics[width=80mm,height=80mm,keepaspectratio]{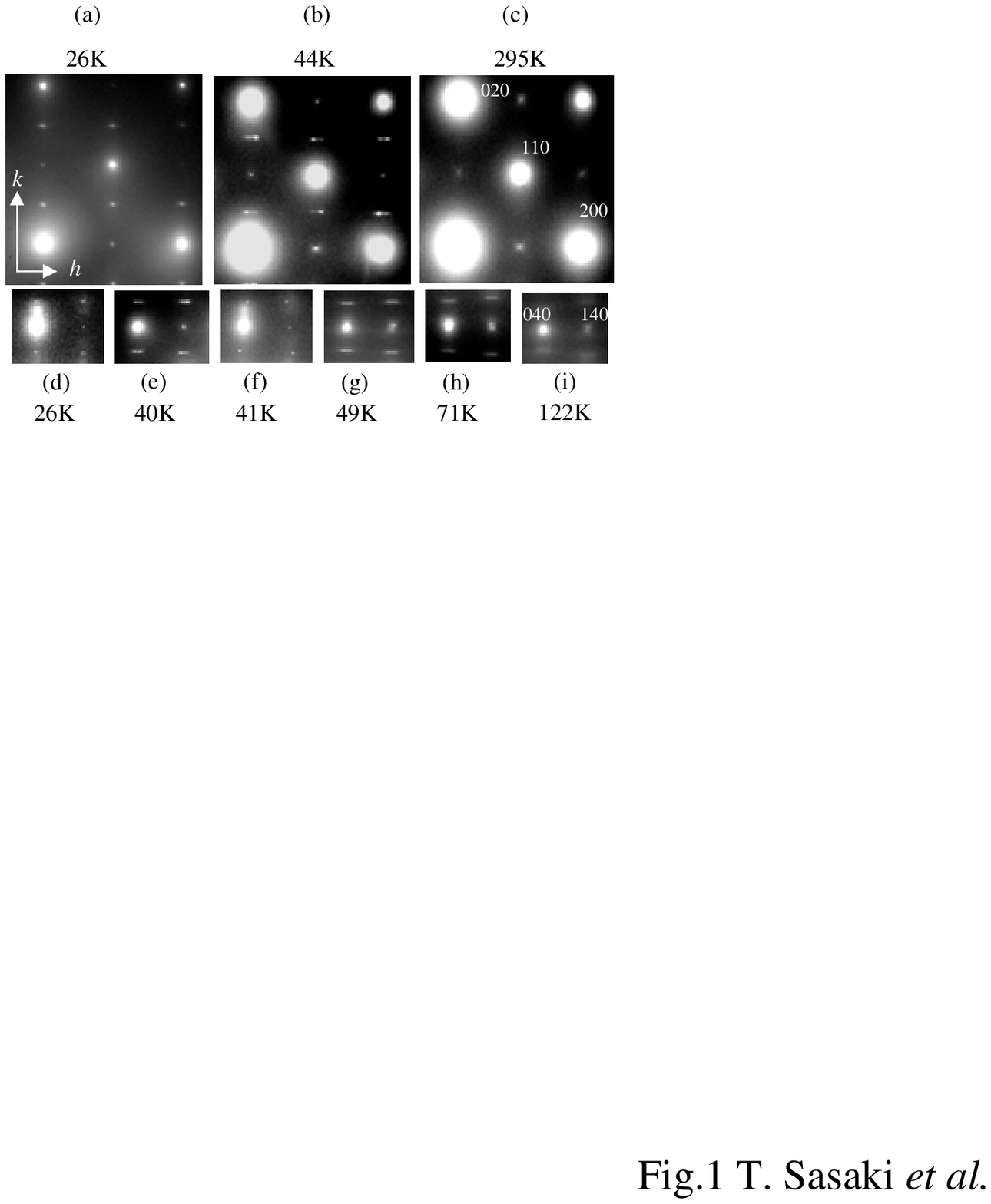}
\caption{}
\end{figure}

\newpage

\begin{figure}[htbp]
\includegraphics[width=90mm,height=90mm,keepaspectratio]{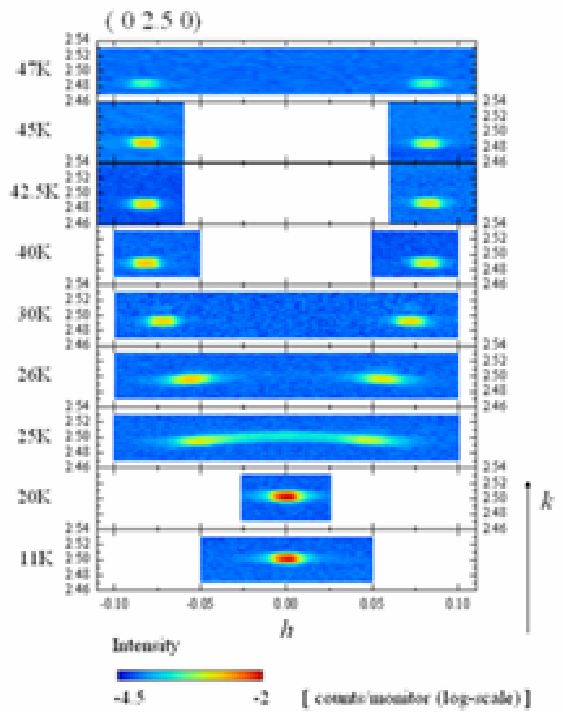}
\caption{}
\end{figure}

\newpage
\begin{figure}[htbp]
\includegraphics{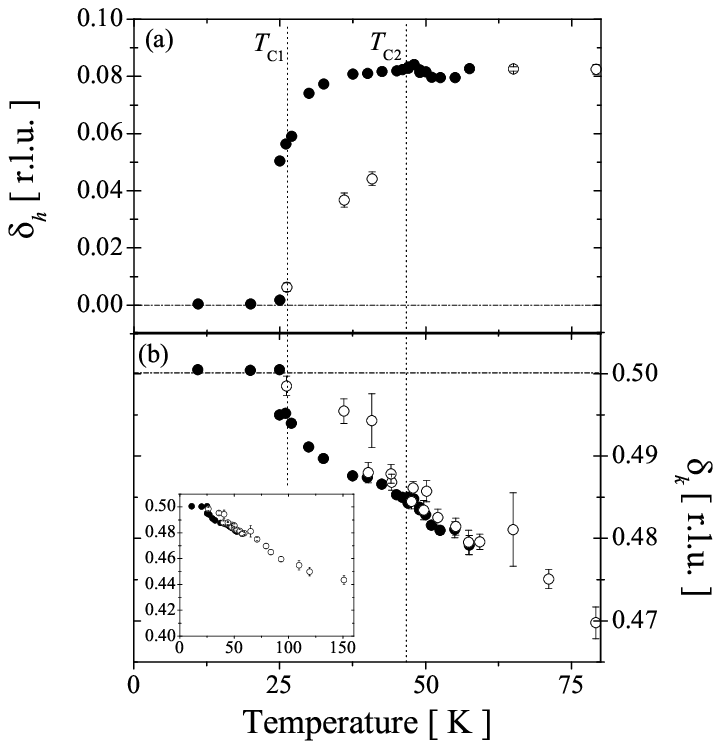}
\caption{}
\end{figure}
\newpage
\begin{figure}[htbp]
\includegraphics[width=80mm,height=80mm,keepaspectratio]{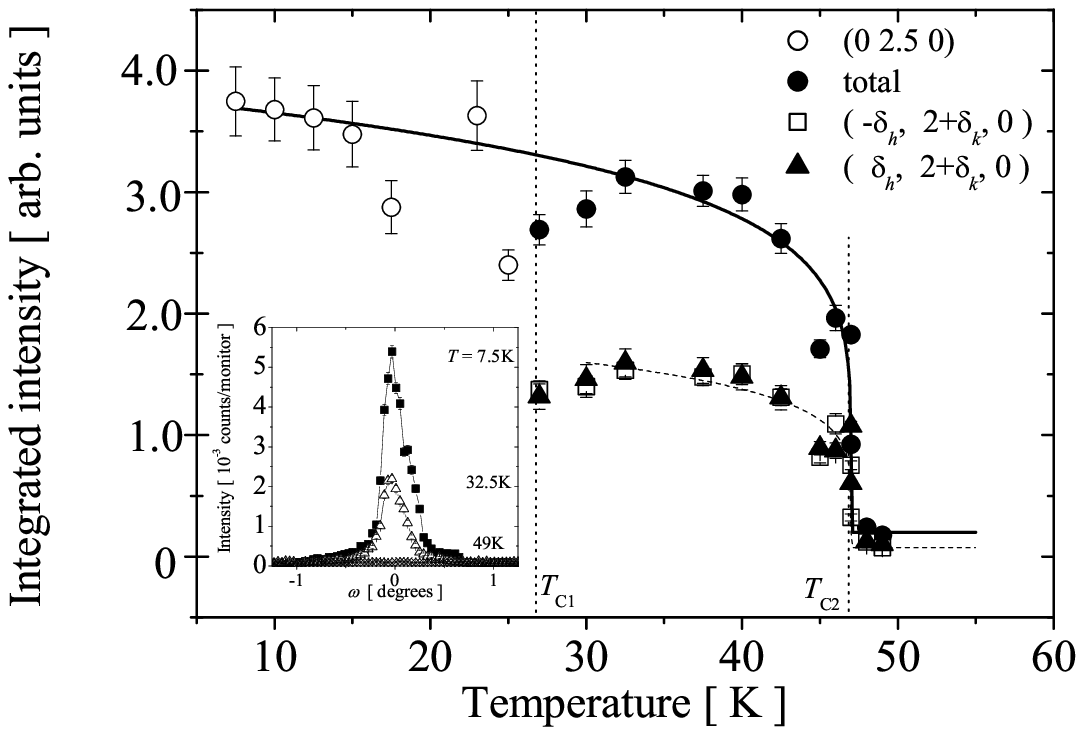}
\caption{}
\end{figure}
\newpage
\begin{figure}[htbp]
\includegraphics[width=80mm,height=80mm,keepaspectratio]{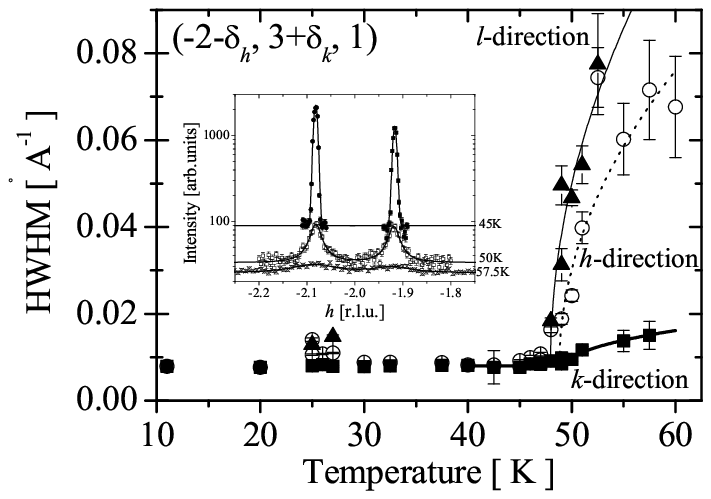}
\caption{}
\end{figure}


\begin{thebibliography}{99}
\bibitem{Comment} We attempt to measure the integrated intensity of each incommensurate reflection several times. Actually, the values at ($\pm \delta_h$, $2+\delta_k$, 0) indicate the difference in intensity, although each intensity indicates the same behavior with decreasing temperature. Because the incommensurate structure has two domains, the intensity represents the ratio of the domains illuminated by the probes.
\bibitem{fit1} The data from 40-48 K is fitted with $T_{\textrm c2}=47.8 \pm 1.0$K and $2\beta=0.513 \pm 0.186$, although Figure 4 does not show the line.
\end{thebibliography}
\end{document}